\documentclass[aps,pre,twocolumn,groupedaddress,superscriptaddress,showpacs]{revtex4}

\usepackage{graphicx}
\usepackage{color}
\usepackage{amsmath}
\usepackage{amsfonts}
\usepackage{amssymb}
\usepackage{dcolumn}
\usepackage{float}

\begin{document}
  \title{q-state Potts model on the Apollonian network}

  \author{Nuno A. M. Ara\'ujo}
    \email{nuno@ethz.ch}
    \affiliation{Computational Physics for Engineering Materials, IfB, ETH Zurich, Schafmattstr. 6, 8093 Zurich, Switzerland}

  \author{Roberto F. S. Andrade}
    \email{randrade@ufba.br}
    \affiliation{Instituto de F\'isica, Universidade Federal da Bahia, 40210-210 Salvador, Brazil}

  \author{Hans J. Herrmann}
    \email{hans@ifb.baug.ethz.ch}
    \affiliation{Computational Physics for Engineering Materials, IfB, ETH Zurich, Schafmattstr. 6, 8093 Zurich, Switzerland}
    \affiliation{Departamento de F\'isica, Universidade Federal do Cear\'a, Campus do Pici, 60451-970 Fortaleza, Cear\'a, Brazil}


  \begin{abstract}
    The q-state Potts model is studied on the Apollonian network with Monte Carlo simulations and the Transfer Matrix method.
    The spontaneous magnetization, correlation length, entropy, and specific heat are analyzed as a function of temperature for different number of states, $q$.
    Different scaling functions in temperature and $q$ are proposed. 
    A quantitative agreement is found between results from both methods.
    No critical behavior is observed in the thermodynamic limit for any number of states.
  \end{abstract}

  \maketitle

  \section{Introduction}\label{sec::intro}

    In the last five years, the Apollonian network (AN) \cite{Andrade05} has attracted a lot of attention from the community working on models on non-Euclidean lattices.
    In particular complex networks \cite{Boccaletti,CostaAdPhys,Dorogo2008} found widespread use in the investigation of most diverse scientific topics, as they can be able to represent connections between individual degrees of freedom in many complex systems.
    Also the behavior of magnetic models on random complex networks has been investigated \cite{Dorogo2002,Aleksiejuka02}.
    AN's have the appealing advantages of geometrical sets defined through an exact inflation rule, where renormalization techniques, leading to exact results can be applied \cite{Migdal75,Berker79,Andrade93}. 
    In fact, this type of techniques has been considered for different hierarchical structures to study critical phenomena \cite{Ravasz03,Hinczewski06,Radicchi08,Boettcher09,deSimoi09}.
    This important property has also motivated using AN, as a first approximation, to study problems from several different areas, such as physiology, geology, communication, energy, and fluid transport \cite{Pellegrini,Vieira2007,Adler1985,Moreira06,Lind07,Schwammle07,Auto08,Kaplan09,Oliveira09,Oliveira10,Oliveira2010}.

\begin{figure}
      \includegraphics[width=0.7\columnwidth]{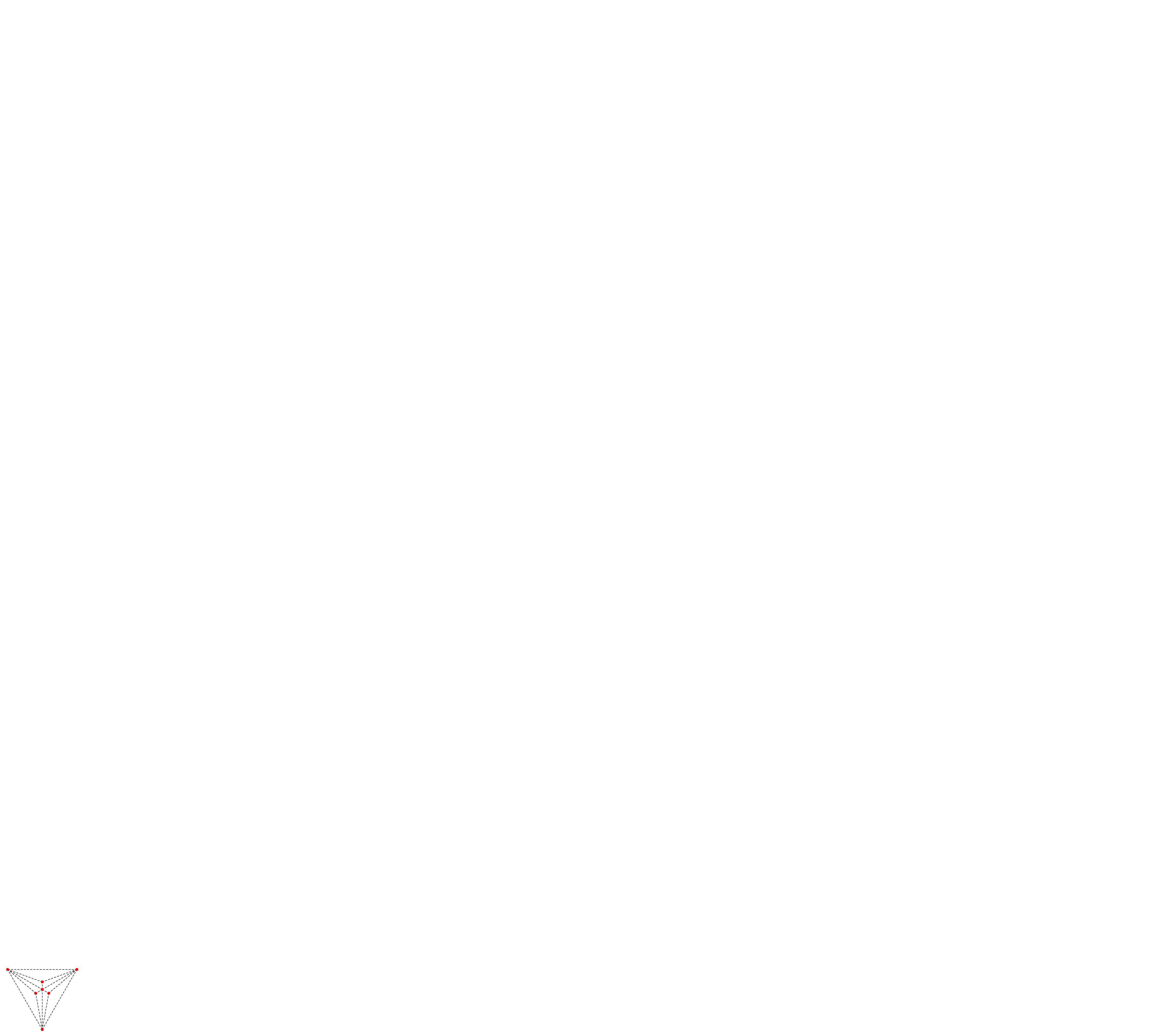}
      \caption{ Apollonian network of generation two.
        \label{fig::apollonian}
      }
    \end{figure}

    The AN geometrical construction can be obtained recursively by initially taking three nodes in the vertices of an equilateral triangle, as shown in Fig.~\ref{fig::apollonian}.
    A new node is then inserted in the center linked to those three.
    Sequentially, new nodes are included, linked to each set of three connected nodes \cite{Andrade05,Doye05}.
    The resulting network is scale free (power-law distribution of node degrees) and satisfies basic features of small-world networks, like large clustering coefficient and average minimal path $\ell \sim lnN$, where $N$ represents the number of nodes in the network.

    In this work we concentrate on the properties of ferromagnetic Potts model on AN's.
    The Potts model \cite{Wu82,Tsallis96} represents a natural extension of the binary Ising model, where each spin variable is allowed to occupy a larger number $q$ of independent states.
    This well known feature is sufficient to produce a richer spectrum of phenomena as compared to the binary Ising model, as the influence of $q$ on the properties of the phase transitions for models defined on Euclidean lattices.
    In turn, this raises the immediate question: how the Potts model, for $q>2$, differs from the Ising model ($q=2$) on AN's, where no phase transition at finite temperature has been detected.
    Does an increase in the value of $q$ lead to a different scenario?

    The investigation of Ising models has already considered a larger number of different situations.
    These include the next-neighbor interaction, which can be of ferro- or antiferromagnetic nature, coupling constants depending upon the generation where they were introduced into the model, or on the number of neighbors of each site \cite{Andrade05b,Andrade09}.
    Also the effect of quenched disorder on the behavior of Ising variables has been analyzed \cite{Kaplan09}.
    In this work, however, we consider only the simplest situation, i.e., homogeneous ferromagnetic couplings with a homogeneous magnetic field. 
    Our results are based on the independent use of two techniques: the standard Monte-Carlo (MC) simulations and the transfer matrix (TM) technique to numerically compute the partition function of the system.
    In both situations, we analyzed how the results depend on the size of the network.
    MC simulations consider up to $9844$ lattice sites. 
    On the other hand, TM technique allows to go to larger sizes.
    In fact, the size can be chosen always sufficiently high to require a numerical convergence of the thermodynamical functions.
    As we show here, quite good agreement between the results from the two different techniques can been achieved. 

    The paper is organized as follows: Section~\ref{sec::apo.potts} introduces the basic properties of AN networks and of the Potts model.
    Section~\ref{sec::tm} brings some details of the used TM method. 
    Section~\ref{sec::res} is divided into two subsections where we discuss, separately, the results obtained from MC simulations and TM maps.
    Finally, in Section~\ref{sec::conc} we present our concluding remarks.

  \section{Apollonian network and Potts model}\label{sec::apo.potts}

    The simplest Apollonian network (AN) \cite{Andrade05,Doye05} is obtained recursively by first placing three nodes at the corner of a triangle (generation $0$).
    A new site is put into the triangle and connected to all three corner nodes forming three new triangles (generation $1$).
    Then, at each generation, a node is placed into each triangle and connected with its three corner nodes.
    Being $n$ the generation, the number of nodes, $N$, is given by
      \begin{equation}\label{eq::num.nodes.gen}
        N(n)= \frac{3^n+5}{2} \ \ ,
      \end{equation}
    \noindent and the number of edges, $E$, by
      \begin{equation}\label{eq::num.edges.gen}
        E(n)= \frac{3^{n+1}+3}{2} \ \ .
      \end{equation}
    \noindent In the limit of large $n$, it is straightforward that $E(n)/N(n)\rightarrow 3$, i.e., on average each node is linked with six other nodes.
    The distribution of links is heterogeneous and the network is scale free, with a degree exponent $\gamma\approx1.585$ \cite{Andrade05}.

    In the q-state Potts model, each node of the network contains a spin which can assume $q$ different states, $\sigma$, i.e., $\sigma=1,2,3,...,q$.
    The Hamiltonian of the model is then,
      \begin{equation}\label{eq::hamiltonian}
        \mathcal{H}= -\sum_{ij}J_{ij}\delta\left(\sigma_i,\sigma_j\right)-h\sum_i\sigma_i \ \ ,
      \end{equation}
    \noindent where the sum runs over directly connected pairs $ij$, yielding nearest-neighbor interactions, and the delta function, $\delta\left(\sigma_i,\sigma_j\right)$, is unity when $i$ and $j$ are in the same state and zero otherwise.
    $J_{ij}$ is the coupling constant which, for simplicity, we consider to be the same for all interconnected pairs, $J_{ij}=J$, and we take the limit $J/k_B=1$, where $k_B$ is the Boltzmann constant.
    We use here the language of spins but, in fact, the q-state Potts model can also be applied to gauge theory, biological patterns, opinion dynamics, and image processing \cite{Wu82,Bentrem10,Araujo10}.
    More recently, a generalized version of the model has been proposed to study the topology of networks through the identification of different communities as well as the overlap between them \cite{Reichardt04,Reichardt06,Fortunato10}.
    In the next section we discuss how to use a transfer matrix formalism to study the q-state Potts model on the Apollonian network.

  \section{Transfer matrix and recurrence maps}\label{sec::tm}

    We have used a transfer matrix (TM) formalism to numerically evaluate the partition function for several Ising models on Apollonian networks \cite{Andrade05b,Andrade09}. 
    This is a very useful method as it yields the properties of the system for any given generation $n$. 
    It is also possible to reach numerically the thermodynamic limit, where the free energy per spin and its derivatives can be obtained to any pre-established precision (usually $\leq 10^{-12}$). 
    The method takes advantage from the AN scale invariance when we go from generation $n$ to $n+1$, and from the fact that partial sums over all spins at generation $n$ can be recursively performed when we write the partition function for generation $n+1$.

    Of course the same strategy can also be used when Potts spins are considered, the only difference being that we must consider matrices of order $q\times q$ and $q\times q^2$.
    For the sake of simplicity, let us write down explicitly the case $q=3$, with $h=0$. 
    The generation zero, $n=0$, consists only of the three spins placed at the vertices of the largest triangle in Fig.~\ref{fig::apollonian}.
    If we perform a partial trace over the spin on the lower vertex, the interaction between the sites $i$ and $k$ can be condensed in a single TM as:

  \begin{widetext}
  \begin{equation}\label{eq3}M_0=
    \begin{pmatrix}
      a_{0} & b_{0} & b_{0} \\
      b_{0} & a_{0} & b_{0} \\
      b_{0} & b_{0} & a_{0}
    \end{pmatrix}=
    \begin{pmatrix}
      a(a^{2}+(q-1)b^{2}) & b(2ab+(q-2)b^2) & b(2ab+(q-2)b^2) \\
      b(2ab+(q-2)b^2) & a(a^{2}+(q-1)b^{2}) & b(2ab+(q-2)b^2) \\
      b(2ab+(q-2)b^2) & b(2ab+(q-2)b^2) & a(a^{2}+(q-1)b^{2})
    \end{pmatrix},
  \end{equation}
  \end{widetext}

    \noindent where $a=\exp(\beta J)$ and $b=1$. 
    To proceed further with the method and consider generation $n=1$, it is necessary to define a $q\times q^2$ TM $L_0$, which describes the interactions among sites $i$, $j$, and $k$.
    We use a column label $\kappa$ that depends on the pair $(j,k)$ according to the lexicographic order, i.e., $\kappa=q(j-1)+k$, so that
    \begin{eqnarray}\label{eq4}L_0=
      \begin{pmatrix}
        r_{0} & s_{0} & s_{0} & s_{0} & s_{0} & t_{0} & s_{0} & t_{0} & s_{0} \\
        s_{0} & s_{0} & t_{0} & s_{0} & r_{0} & s_{0} & t_{0} & s_{0} & s_{0} \\
        s_{0} & t_{0} & s_{0} & t_{0} & s_{0} & s_{0} & s_{0} & s_{0} & r_{0}
      \end{pmatrix} 
      =\\ 
      \begin{pmatrix}
        a^{3} & ab^{2} & ab^{2} & ab^{2} & ab^{2} & b^{3} & ab^{2} & b^{3} & ab^{2} \\
        ab^{2} & ab^{2} & b^{3} & ab^{2} & a^{3} & ab^{2} & b^{3} & ab^{2} & ab^{2} \\
        ab^{2} & b^{3} & ab^{2} & b^{3} & ab^{2} & ab^{2} & ab^{2} & ab^{2} & a^{3}
      \end{pmatrix}. \nonumber
    \end{eqnarray}

    As discussed in detail in Ref.~\cite{Andrade05}, transfer matrices $M_1$ and $L_1$ can be expressed in terms of $M_0$ and $L_0$ as 

    \begin{equation}\label{eq5}
      (M_{1})_{i,k} = \sum_{j=1}^q\sum_{\ell=1}^q(L_0)_{i,j\ell} (L_0)_{i,\ell k} (L_0^t)_{k,j\ell} \ \ ,
    \end{equation}

    \noindent and

    \begin{equation}\label{eq6}
      (L_{1})_{i,jk} = \sum_{\ell=1}^q(L_0)_{i,j\ell} (L_0)_{i,\ell k}(L_0^t)_{k,j\ell} \ \ .
    \end{equation}

    Note that the above matrix maps require that the matrix element $t_0$ ($t_n$) needs to be introduced only for integer $q>2$. 
    The matrix elements $a_n$ and $b_n$ can be expressed as $a_n=r_n+(q-1)s_n$ and $b_n=2s_n+(q-2)t_n$.

    Since the network grows according to a generation independent inflation rule, Eqs.~(\ref{eq5})~and~(\ref{eq6}) apply for any value of $n$, just by replacing $1$ by $n+1$ and $0$ by $n$. Such general matrix maps can be rewritten in terms of maps for the matrix elements of $L_{n+1}$ in terms of those of $L_{n}$ as
    \begin{equation}\label{eq7}
      \begin{array}{l}
        r_{n+1} = r_n^3 + (q-1)s_n^3\\
        s_{n+1} = r_ns_n^2 + s_{n}^3 + (q-2)s_nt_n^2\\
        t_{n+1} = 3s_n^2t_n + (q-3)t_{n}^3
      \end{array} \ \ ,
    \end{equation}

    \noindent from which the elements $a_{n+1}$ and $b_{n+1}$ can be obtained. 
    It is possible to directly evaluate the free energy and all other thermodynamic functions in terms of $a_{n}$ and $b_{n}$ or, equivalently, from $r_{n}$, $s_{n}$, and $t_{n}$. 
    Due to the $M_n$'s particular form, its eigenvalues can be easily evaluated as $\Lambda_n=a_n+(q-1)b_n$ and the ($q\!\!-\!\!1$)-degenerated $\lambda_n=a_n-b_n$. 
    Since the numerical values of $a_n, b_n, r_n, s_n,$ and $t_n$ grow exponentially, it is convenient to write down recurrence maps that avoid numerical overflows when they are iterated. 
    Since $r_n$ is the fastest exponentially growing variable, this can be accomplished by deriving maps for the free energy $f_n=-T\ln(\Lambda_n)/N(g)$, the correlation length $\xi_n=1/\ln(\Lambda_n/\lambda_n)$, and the auxiliary variable $y_n=t_n/r_n$. 
    An alternative definition would be to replace the map for $\xi_n$ by that for $x_n=s_n/r_n$, from which the value of $\xi_n$ can be evaluated. 
    This set of maps can be enlarged by working out explicit recurrence relations for the derivatives of $f_n(T)$ and the other variables ($y_n$, $x_n$, or $\xi_n$) with respect to the temperature. 
    This way, the explicit temperature dependence of the entropy $s(T)$ and the specific heat $c(T)$ can be obtained.

    To evaluate the magnetic properties, it is necessary to break the symmetry among the $q$ states and insert a non zero field $h\neq0$ along one of the $\sigma$ (say $\sigma=1$) directions.
    This changes the simple structure of matrices $M_n$ and $L_n$, that then have a much larger number of matrix elements. 
    By performing an explicit calculation of the matrix elements for larger values of $q$, it is possible to complete the set of maps used in this work, as listed in the Appendix.

  \section{Results and discussion}\label{sec::res}

    To study the q-state Potts model on the Apollonian network we consider two different approaches: Monte Carlo simulations (MC) and the Transfer Matrix method (TM).
    Monte Carlo simulations are limited to small system sizes, so results are affected by finite-size effects.
    On the other hand, with the Transfer Matrix formalism large system sizes can be considered allowing to compute the thermodynamic and magnetic properties, numerically, in the thermodynamic limit.
    This section is then divided in two subsections.
    In the first we discuss the results obtained with Monte Carlo simulations for the magnetization, $m(q,T)$, and the specific heat, $c(q,T)$, as a function of the temperature, $T$, and the number of states, $q$.
    A comparison between MC and TM results is also included.
    In the second part, not only the magnetization and specific heat but also the entropy, $s(q,T)$, and the correlation length, $\xi(q,T)$, are obtained with the TM technique for larger system sizes.
    The thermodynamic limit is then discussed.

    \subsection{Monte Carlo simulations}\label{subsec::mc}

      \begin{figure}
        \includegraphics[width=\columnwidth]{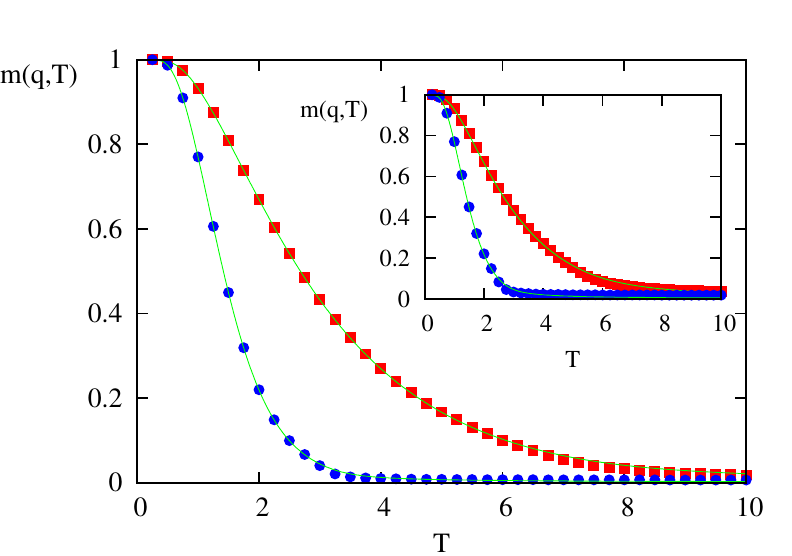}
        \caption{
            (Color online) Magnetization, $m(q,T)$, as a function of the temperature, $T$, for generation $7$ (inset) and $9$ (main plot).
            The considered number of states, $q$, are $2$ (red squares) and $8$ (blue circles).
            The lines were obtained with Transfer Matrices and the points with Monte Carlo simulations.
            Simulational results were averaged over $10^4$ samples.
            Error bars are within the point size.
          \label{fig::mcmagT}
        }
      \end{figure}

      We carried out Monte Carlo simulations for two different generations of the Apollonian network, $7$ and $9$, corresponding, respectively, to $1096$ and $9844$ spins (Eq.~(\ref{eq::num.nodes.gen})).
      For each system size, we considered two values of $q$ (number of states), $2$ and $3$, and we computed the magnetization and specific heat as a function of the temperature.
      For the Potts model \cite{Binder81}, the magnetization is defined as
        \begin{equation}\label{eq::def.mag}
          m(q,T)= \frac{q(n_1(q,T)-1)}{q-1} \ \ ,
        \end{equation}
      \noindent where $n_1$ is the fraction of spins in the state $1$, and the specific heat is defined as
        \begin{equation}\label{eq::def.specheat}
          c(q,T)= \frac{1}{N}\left(\frac{J}{k_BT}\right)^2\left[<u^2>-<u>^2\right] \ \ ,
        \end{equation}
     \noindent where $N$ is the total number of spins and $<u>$ and $<u^2>$ are, respectively, the first and second moment of the energy per spin.
      All Monte Carlo results were averaged over $10^4$ samples.

      In Fig.~\ref{fig::mcmagT} we show the magnetization, $m(q,T)$, as a function of the temperature for $q=2$ and $q=8$, for generation $9$.
      The inset shows the same functions for generation $7$.
      The points correspond to Monte Carlo results and the lines to numerical results obtained with the Transfer Matrix method.
      An exponential decrease of the magnetization is observed.
      The larger the number of possible states the steeper the change with temperature.
      We find an agreement between the Monte Carlo results and the ones from the Transfer Matrix method.

      \begin{figure}
        \includegraphics[width=\columnwidth]{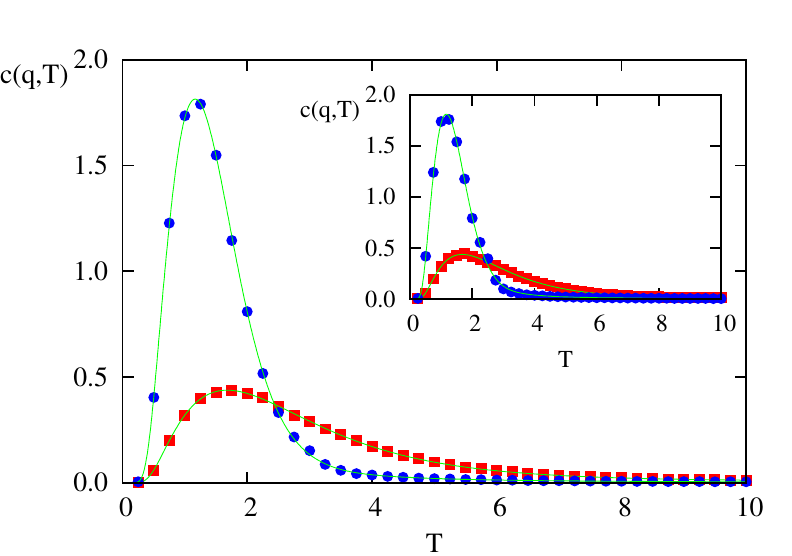}
        \caption{
            (Color online) Specific heat, $c(q,T)$, as a function of the temperature, $T$, for generation $7$ (inset) and $9$ (main plot).
            The considered number of states, $q$, are $2$ (red squares) and $8$ (blue circles).
            The lines were obtained with Transfer Matrices and the points with Monte Carlo simulations.
            Simulational results were averaged over $10^4$ samples.
            Error bars are within the point size.
          \label{fig::mcspechT}
        }
      \end{figure}

      Figure~\ref{fig::mcspechT} shows the specific heat, $c(q,T)$, as a function of temperature.
      Like in Fig.~\ref{fig::mcmagT}, two different values of states ($2$ and $8$) and generations ($7$ and $9$) are considered.
      For both values of $q$ a Schottky maximum in the specific heat is found for values of $T$ below $2$.
      With increasing number of states the maximum becomes sharper and occurs at lower temperature (discussed in the next subsection).
      Results for different generations reveal no significant size effects in the specific heat.
      A quantitative agreement between the Monte Carlo (points) and Transfer Matrices (lines) results is obtained.

      Results from Monte Carlo simulations are affected by finite-size effects.
      In the next subsection, we consider the Transfer Matrix formalism to study larger system sizes and discuss the behavior of the system in the thermodynamic limit.

    \subsection{Transfer Matrix}\label{subsec::tm}
 
      \begin{figure}
        \includegraphics[width=\columnwidth]{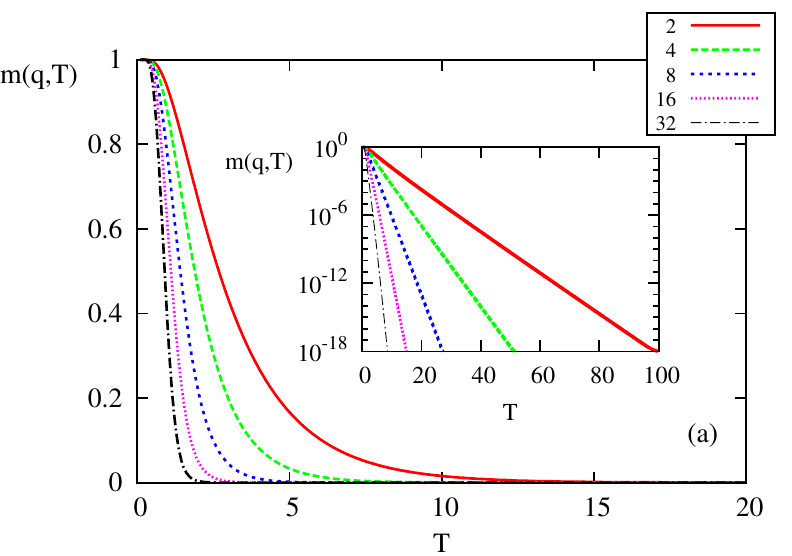}
        \includegraphics[width=\columnwidth]{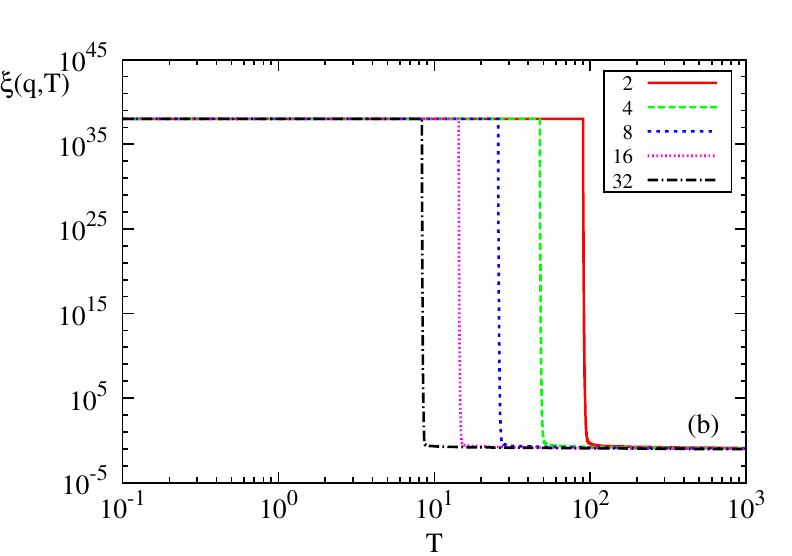}
        \caption{ 
          (Color online) a) Magnetization, $m(q,T)$ as a function of the temperature, $T$, for $q=\{2,4,8,16,32\}$ (from right to left).
          The magnetization vanishes exponentially with the temperature (see inset) and the argument of the exponential decreases linearly with $q$.
          b) Correlation length, $\xi(q,T)$, as a function of the temperature, $T$, for the same values of $q$.
          Results obtained for generation $100$.
          \label{fig::magT}
        }
      \end{figure}

     We evaluated the partition function for the q-state Potts model on an Apollonian network by numerically iterating the transfer matrix maps introduced in Sec.~\ref{sec::tm} and detailed in the Appendix.
     We study thermodynamic and magnetic properties like the spontaneous magnetization, $m$, the specific heat, $c$, the entropy, $s$, and the correlation length, $\xi$, for different values of $q$, as a function of the temperature, $T$.

     Figure~\ref{fig::magT}(a) shows the magnetization, $m(q,T)$, as a function of the temperature, $T$, for different values of the number of states, $q=\{2,4,8,16,32\}$.
     A smooth decay of the magnetization with the temperature is obtained,
       \begin{equation}\label{eq::mag.q.T}
         m(T,q)\sim \exp\left(-\phi(q) T\right) \ \ ,
       \end{equation}
     \noindent where $\phi(q)$ is a linear function of $q$,
       \begin{equation}\label{eq::phi.mag.q}
         \phi(q)= aq+b \ \ ,
       \end{equation}
     \noindent with units of the inverse of temperature.
     We have estimated $a=0.18\pm0.01$ and $b=0.17\pm0.03$, and no significant finite-size effects are observed.
     For $q=2$, the q-state Potts model is equivalent to the Ising model and we recover the previously reported behavior for this system \cite{Andrade05b}.
     Note that, the temperature in the Ising model, $T_{Ising}$, differs from the one of the Potts model, $T_{Potts}$, by a factor of $1/2$, i.e., $T_{Ising}=T_{Potts}/2$.
     Consequently, a decay of the magnetization, in the Ising model, as $\exp(-T)$, corresponds to $\exp(-T/2)$ here.

    \begin{figure}
      \includegraphics[width=\columnwidth]{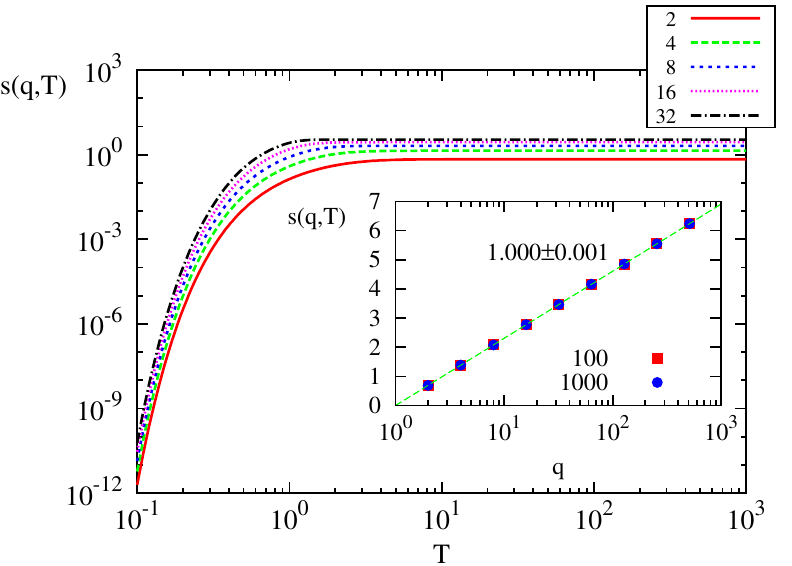}
      \caption{ 
        (Color online) Entropy, $s(q,T)$, as a function of the temperature, $T$, for $q=\{2,4,8,16,32\}$ (from bottom to top).
        Results obtained for generation $100$.
        For all values of $q$ the entropy converges to a fixed value for high temperatures.
        In the inset we see this value as a function of $q$ at two different temperatures: $100$ (red squares) and $1000$ (blue dots).
        Above a certain temperature, the entropy scales logarithmically with $q$.
        \label{fig::entropyT}
      }
    \end{figure}

     The behavior of the magnetization shows no evidence of a sharp order-disorder transition, which is not the case for the correlation length, $\xi(q,T)$, in Fig.~\ref{fig::magT}(b).
     For all values of $q$ there is a well-defined temperature, $T^*$, below which the correlation length numerically diverges when compared with the value of $\xi$ in the disordered phase.
     Here we chose the divergence threshold to be $10^{38}$, but any other value would lead to the same qualitative picture.
     Above $T^*$ the correlation length attains finite values.
     This apparent transition is in fact a finite-size effect.
     As observed for the Ising model \cite{Andrade05b}, the value of $T^*$ goes linearly with the generation, $n$, being infinite in the thermodynamic limit, i.e., $T^*\sim\theta(q)n$, where, for large $q$,
     \begin{equation}\label{eq::theta}
       \theta(q)\sim\log^\eta(q) 
     \end{equation}
     \noindent and $\eta= -3.81\pm0.07$.
     The value of $T^*$ decreases with $q$ according to,
     \begin{equation}\label{eq::tstarq}
       T^*\sim\log^\xi(q) \ \ ,
     \end{equation}
     \noindent with $\xi= -2.74\pm0.04$.
     The absence of criticality is in agreement with previous results, where several models characterized by criticality on periodic networks show no critical transition, for finite temperature, in scale-free networks with a degree exponent $\gamma\leq3$ \cite{Igloi02,Goltsev03,Kwak07,Karsai07}.
     In this regime, an ordered phase is observed at any temperature.
     Kaplan, Hinczewski, and Berker \cite{Kaplan09} reported that in the presence of quenched disorder the ordered phases are still robust and persist for the entire range of disorder.
     Recently, Igl\'oi and Turban \cite{Igloi02} have considered a mean-field version of the Potts model and reported that an order/disorder transition solely occurs for $\gamma>3$.
     A $q(\gamma)$ can then be defined above which the order of the transition changes from second- to first order, like on periodic lattices.

      \begin{figure}
        \includegraphics[width=\columnwidth]{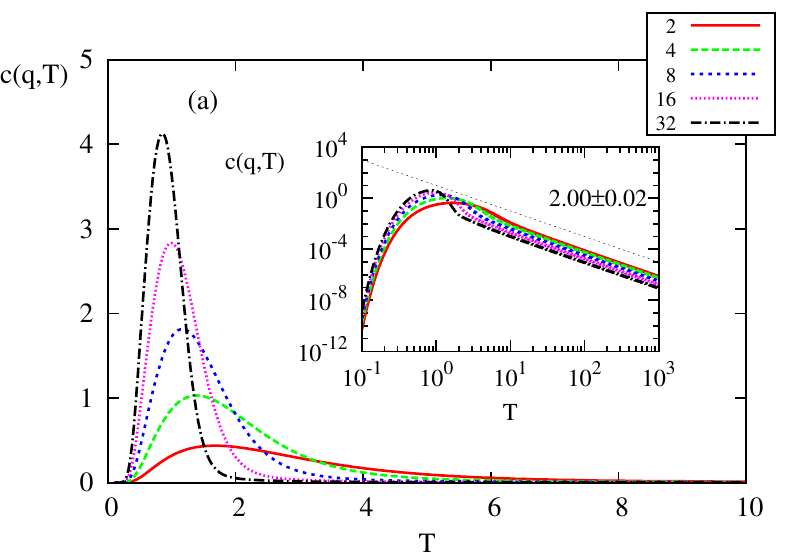}
        \includegraphics[width=\columnwidth]{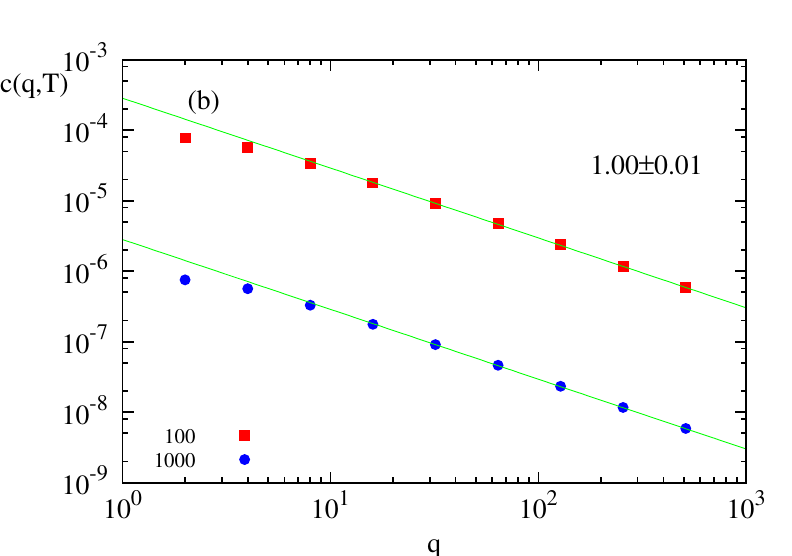}
        \caption{ 
          (Color online) (a) Specific heat, $c(q,T)$, as a function of the temperature, $T$, for $q=\{2,4,8,16,32\}$ (from bottom to top).
          Results are obtained for generation $100$.
          For high temperatures, the specific heat scales with the temperature according to a power law (inset in (a)).
          (b) Specific heat as a function of $q$ at two temperatures: $100$ (red squares) and $1000$ (blue dots).
          \label{fig::specheatT}
        }
      \end{figure}

     In Fig.~\ref{fig::entropyT} we draw the curves of the entropy, $s(q,T)$, as a function of the temperature.
     For all considered values of $q$, the entropy rises with $T$ and saturates at a maximum for large temperatures (above $10$).
     The inset of Fig.~\ref{fig::entropyT} contains the entropy as a function of the number of states, $q$, for two different temperatures, $100$ and $1000$.
     For large $T$, the entropy depends logarithmically on $q$,
       \begin{equation}\label{eq::entropy.q}
         s(q,T)\sim \log(q) \ \ .
       \end{equation}
     \noindent These results have no dependence on the system size.
     As for the specific heat (discussed below), the curve of the entropy converges fast with the generation number.

     The specific heat, $c(q,T)$, is shown in Fig.~\ref{fig::specheatT}(a).
     As already observed for smaller system sizes with Monte Carlo simulations, the dependence of the specific heat on the temperature is smooth with a peak at a temperature below $2$.
     No divergence of the maximum with the system size is observed as would be expected for a transition.
     In fact, this peak corresponds to a Schottky maximum and is due to a fast increase in the entropy (see Fig.~\ref{fig::entropyT}).
     The larger the value of $q$ the sharper the peak and the lower the temperature at which it occurs.
     For large $T$ the specific heat diminishes with temperature according to a power law with an exponent $2.00\pm0.02$.
     This exponent is the same for different values of $q$ and the behavior is independent of the system size.
     In this power-law regime, for a fixed temperature, the specific heat scales linearly with $q$ (Fig.\ref{fig::specheatT}(b)).
     Therefore, for large $T$, the following scaling law can be postulated,
       \begin{equation}\label{eq::scale.spech}
         c(q,T)\sim 1/q T^{-2} \ \ .
       \end{equation}

  \section{Conclusions}\label{sec::conc}

    We studied the q-state Potts model on the Apollonian network through Monte Carlo simulations and Transfer Matrix method.
    Different scaling relations were obtained for magnetic and thermodynamic properties like the spontaneous magnetization, correlation length, entropy, and specific heat.
    We have shown that the magnetization decays smoothly with the temperature and the greater the number of states the steeper the decay.
    However, no order-disorder transition is found in the thermodynamic limit for any value of the number of states, $q$.
    The specific heat is characterized by a Schottky maximum which becomes sharper with increasing $q$, without divergence as expected for a transition.
    As previously reported for the Ising model on the same network \cite{Andrade05b,Andrade09}, the specific heat converges rapidly with the generation number to the thermodynamic limit.

    In the present work we consider spins interacting ferromagnetically with its nearest neighbors in the absence of a magnetic field.
    A more general version of the model can be considered where pairs of spins interact with further neighbors, e.g., next-nearest neighbors, and a magnetic field, either uniform or site dependent.
    Besides, it would be interesting to analyze the effect of different coupling constants between spins based on their generation \cite{Andrade09}, to observe possible occurrence of a critical behavior in the thermodynamic and magnetic properties when the temperature changes. 
    The considered methodology can also be taken to study the properties of the recently proposed versions of the model to identify communities in networks \cite{Reichardt04,Reichardt06,Fortunato10}.
    
  \section{Appendix}

    The map for the free energy assumes a simpler form if we consider an alternative definition $\overline{f}=-T\ln(r_n)/N(g)$, from which the value $f$ can be easily obtained. 
    In fact, in the limit of large $n$, $f$ and $\overline{f}$ become identical.
    The full set of maps, which depends also on the variable $g=exp(\beta h)$ with $\beta=1/T$, reads:

    \begin{equation}
      \label{A1}
      \overline{f}_{n+1}=\frac{3N_n \overline{f}_n}{N_{n+1}}-
      \frac{T}{N_{n+1}}\ln (g + (q-1)v_n^3)
    \end{equation}

    \begin{equation}
      \label{A2}
      u_{n+1} =\frac{\overline{r}_{n+1}}{r_{n+1}}=\frac{u_n^3+gw_n^3+(q-2)x_n^3}{g + (q-1)v_n^3}
    \end{equation}

    \begin{equation}
      \label{A3}
      v_{n+1} =\frac{s_{n+1}}{r_{n+1}}=\frac{gv_n^2+v_nw_n^2+(q-2)v_ny_n^2}{g + (q-1)v_n^3}
    \end{equation}

    \begin{equation}
      \label{A4}
      w_{n+1} =\frac{\overline{s}_{n+1}}{r_{n+1}}=\frac{gv_n^2w_n+u_nw_n^2+(q-2)x_ny_n^2}{g + (q-1)v_n^3}
    \end{equation}

    \begin{equation}
      \label{A5}
      x_{n+1} =\frac{\hat{s}_{n+1}}{r_{n+1}}=\frac{gw_ny_n^2+u_nx_n^2+x_n^3+(q-3)x_nz_n^2}{g + (q-1)v_n^3}
    \end{equation}

    \begin{equation}
      \label{A6}
      y_{n+1} =\frac{t_{n+1}}{r_{n+1}}=\frac{gy_nv_n^2+2w_nx_ny_n+(q-3)y_n^2z_n}{g + (q-1)v_n^3}
    \end{equation}

    \begin{equation}
      \label{A7}
      z_{n+1} =\frac{\overline{t}_{n+1}}{r_{n+1}}=\frac{gy_n^3+3x_n^2z_n+(q-4)z_n^3}{g + (q-1)v_n^3}
    \end{equation}

    New variables $\overline{r}_n,\overline{s}_n,\hat{s}_n,$ and $\overline{t}_n$ have been introduced because the external field $h$ reduces the problem symmetry, as reflected in a larger number of distinct matrix elements. 
    Note that $\overline{r}_1=r_1,\overline{s}_1=\hat{s}_1=s_1,$ and $\overline{t}_1=t_1$. 
    If $h=0$, the number of independent equations is reduced to three.

  \begin{acknowledgments}
    We acknowledge financial support from the ETH Competence Center Coping with Crises in Complex Socio-Economic Systems (CCSS), through ETH Research Grant CH1-01-08-2, from the Brazilian agencies CNPq, FUNCAP, and FAPESB, and from the National Institute of Science and Technology for Complex Systems, Brazil.
  \end{acknowledgments}

\bibliography{text}

\end{document}